\documentclass[12pt]{article}

\usepackage{graphicx}
\usepackage[a4paper,margin=1in]{geometry}
\usepackage{authblk} % for multiple authors/affiliations
\usepackage{lipsum} % dummy text (remove later)
\usepackage{amsthm}
\usepackage{amsmath}
\newtheorem{theorem}{Theorem}
\usepackage{natbib}
\usepackage{algorithm}
\usepackage{algpseudocode}
\usepackage{booktabs}
\usepackage{tabularx}   % for X column
\usepackage{adjustbox}  % optional, for extra safety
\usepackage[margin=1in]{geometry}
\usepackage{enumitem}
\usepackage{comment}
\usepackage{graphicx}
\usepackage{caption}
\usepackage{subcaption}
\newfloat{procedure}{htbp}{lop}
\floatname{procedure}{Procedure}

\usepackage{lineno}
\usepackage{setspace}
\usepackage{newtxtext,newtxmath, bm}

\title{Estimating Negative Income Distributions via Data Fusion with Vine Copula-based Imputation}

\author[1]{Sithara Wijekoon\thanks{sithara.mudiyanselage@hdr.qut.edu.au}}
\author[1]{Helen Thompson}
\author[2]{Summer Wang}
\author[1]{Gentry White}

\affil[1]{School of Mathematical Sciences, Queensland University of Technology, Australia}
\affil[2]{Australian Bureau of Statistics, Australia}
\date{}
\begin{document}
%\linenumbers
\onehalfspacing
%\RaggedRight
\maketitle
\begin{abstract}

Imputation-based data fusion combines datasets by imputing missing variables in one dataset using information from the other. The validity of imputation-based data fusion relies on accurately preserving both the imputed distributions and the underlying multivariate dependence structure across data sources. However, traditional imputation-based data fusion methods often struggle to preserve the dependence structure, particularly when marginal distributions differ or when complex, nonlinear multivariate dependencies exist. Inadequate preservation of these dependencies can result in distorted joint distributions and biased inference in the fused data. To address this challenge, this study introduces a novel imputation-based data fusion framework that utilises conditional sampling from C-vine and D-vine copulas as the imputation mechanism. This approach flexibly models pairwise and higher-order dependencies while accommodating heterogeneous marginal distributions, thereby generating imputations that more accurately maintain the distribution of the multivariate data. The proposed method is validated through a simulation study using survey data and a real-data application that estimates the negative income distribution in survey data using information from administrative tax records. Results indicate that the proposed method outperforms existing approaches in preserving both the imputed distributions and the dependence structure across datasets.

\end{abstract}

\section{Introduction}

Integrating data from multiple sources is a critical strategy for achieving comprehensive insights, since a single source often lacks sufficient coverage or information to support robust conclusions. A widely used approach to data integration is imputation-based data fusion, also known as statistical matching, which integrates information from multiple datasets by transferring data between sources within an imputation framework, using a set of common variables \citep{Moretti2023Improving, Fosdick2016Categorical, Chiara2019Integrated}.

Imputation-based data fusion employs various imputation techniques, which are generally classified into three categories. Nonparametric methods, such as hot-deck matching, do not rely on explicit distributional assumptions. Parametric methods, typically regression-based approaches, assume a specific functional form for the underlying relationships. Mixed methods, including Predictive Mean Matching (PMM) and propensity score techniques, combine both parametric and non-parametric properties to leverage the strengths of each approach \citep{Chiara2019Integrated}. Traditional data fusion methods, however, often struggle to preserve the dependence structure, particularly when marginal distributions differ or when complex, nonlinear multivariate dependencies exist. For instance, hot-deck matching methods match recipient records with missing values to donor records based on distance measures or random allocation within homogeneous subgroups, which may not accurately retain the joint distribution of variables. Regression models are typically specified under the assumption of linearity or specific functional forms, which may not capture complex nonlinear or higher-order dependencies present in the data. PMM exhibits limitations similar to those of regression models, as it is generally specified under the assumption of linearity. These limitations pose a significant challenge for data fusion, where maintaining the underlying dependence structure across variables is essential for generating coherent fused datasets and ensuring valid downstream analysis.

Machine learning methods, including Classification and Regression Trees (CART) and Random Forests (RF), have recently been adopted in imputation-based data fusion due to their flexibility in capturing non-linear dependencies without relying on restrictive parametric assumptions \citep{Emmenegger2023Evaluating}. However, these approaches are primarily designed for predictive accuracy and often lack a coherent probabilistic framework for statistical inference. As a result, the uncertainty associated with the fused data is not inherently defined within a unified probabilistic framework and cannot be naturally propagated through subsequent statistical analyses. This limitation is particularly important in data fusion, where the fused dataset is often used for downstream inference and decision-making, requiring both the dependence structure and the associated uncertainty to be properly represented. Copula-based approaches address this need by explicitly modelling dependence structures within a statistical framework and enabling coherent uncertainty quantification that can be incorporated into subsequent inferential analysis.

Recent studies have demonstrated the effectiveness of copula-based approaches for handling multivariate missing data. In contrast to traditional imputation techniques, which often assume linear relationships, copulas can capture a broad spectrum of dependence structures, including asymmetric and tail dependencies. Vine copulas extend this framework by decomposing high-dimensional dependencies into sequences of bivariate copulas, thereby facilitating the modelling of pairwise relationships. The hierarchical structure renders vine copulas particularly effective for imputing missing values in high-dimensional and heterogeneous datasets, where conventional methods may fail to preserve dependence structures \citep{Hasler2018Vine, Chapon2023Imp,Ene2013CVine}. To our knowledge, copula-based imputation has not yet been explored in data fusion. Addressing this gap, the present study proposes a novel imputation-based data fusion approach using vine copulas that offers a probabilistic framework for generating imputed values from the underlying joint distribution, preserving both the marginal distributions of the imputed data and the complex dependence structures among variables. By explicitly modelling dependencies and enabling natural propagation of uncertainty, the approach enhances the reliability and interpretability of fused data, making them suitable for downstream statistical analysis and decision-making. 

The proposed approach builds on the conditional sampling algorithm for two types of vine copulas, the canonical vine (C-vine) and the drawable vine (D-vine), proposed by \cite{Bevacqua2017Multivariate}. The proposed method improves the quality of fused data by selecting the optimal vine structure that best preserves the underlying dependence structure by minimising the average difference in correlations between imputed and observed variables across datasets. The proposed approach is validated through a simulation study using survey data and is compared with alternative methods. Additionally, the approach is validated in a real-data application to estimate the negative income distribution in survey data, using administrative data as a benchmark for the true distribution.  

The remainder of this paper is structured as follows. Section 2 outlines the theoretical foundation for data fusion and copula modelling. Section 3 details the proposed data fusion approach. Section 4 presents a simulation study, and Section 5 applies the proposed approach to real-world data and compares it with alternative methods. Section 6 evaluates the results, discusses limitations, and suggests directions for future research.

\section{Methodology}

This paper proposes a novel data fusion framework using vine copulas. To provide the necessary foundation for the proposed approach, this section presents a brief overview of the imputation-based data fusion scenario and copula modelling. 

\subsection{Comparison between imputation and imputation-based data fusion}

Data fusion incorporates imputation as part of its framework; however, the two approaches differ fundamentally in their objectives and implementation. Imputation is designed to handle missing data within a single dataset. In this setting, some values of the variables are missing, and the goal is to replace them with plausible estimates to enable complete-data analysis. The imputation procedure relies on observed relationships among variables within the same dataset, and the quality of the imputation is assessed by its ability to predict the true missing values and preserve the statistical properties of the original dataset. In contrast, imputation-based data fusion creates a fused dataset by integrating two or more datasets that contain different variables measured in different units but share a set of common variables. In this context, the variables to be imputed are not missing due to nonresponse; rather, they are structurally unobserved in one dataset because they were never collected. As a result, the fused dataset enables joint analysis of variables that are not jointly observed in reality. In imputation-based data fusion, the goal is to preserve the joint distribution between the imputed and observed variables in order to produce valid inference. Therefore, methods used for data fusion must be capable of accurately capturing and preserving the dependence structure between variables across datasets, rather than simply minimising prediction error.  

Accordingly, the present study employs imputation as a mechanism for data fusion, and the methods' performance is evaluated in terms of their effectiveness in producing valid and reliable fused data, rather than solely as missing-data imputation techniques. 

\subsubsection{Theoretical aspects of imputation-based data fusion}

\begin{figure}[h]
    \centering
    \includegraphics[scale=0.6]{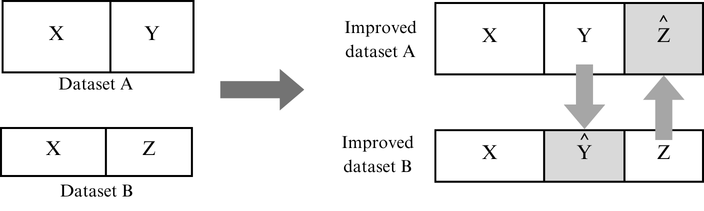}
    \caption{Imputation-based data fusion scenario}
    \label{fig: Imp scenario}
\end{figure}

This section outlines the fundamental framework of imputation-based data fusion. Figure~\ref{fig: Imp scenario} presents the data fusion diagram based on the imputation mechanism. $A$ and $B$ are two datasets that represent the same target population. The variable $Y$, originally from dataset $A$, is not present in dataset $B$, and the variable $Z$, originally from dataset $B$, is not present in dataset $A$. Both datasets share a set of common variables, $X$, which are associated with $Y$ in dataset $A$ and $Z$ in dataset $B$. In this framework, ${Y}$ is imputed into dataset $B$ based on the relationships between the common variables $X$ and the observed $Y$ in dataset $A$, where $B$ is called the recipient file and $A$ is called the donor. As a result, the process yields an improved dataset $B$ comprising all variables $(X,\widehat{Y}, Z)$. Similarly, dataset $A$ can be enriched to contain $(X, Y,\widehat{Z})$ by imputing ${Z}$ into $A$ using the relationships between $X$ and the observed $Z$ in dataset $B$, where $A$ is the recipient and $B$ is the donor. Each improved or fused dataset represents the joint distribution of $(X, Y, Z)$, which cannot be estimated from either data source alone \citep{Chiara2019Integrated, Emmenegger2023Evaluating}. The present study focuses on improving a single dataset (A or B) by using another dataset.

\subsection{Copula functions}
A copula is defined as a multivariate cumulative distribution function in which each marginal distribution follows a uniform distribution on the interval [0,1]. Copulas facilitate the modelling of dependence structures among random variables independently of their marginal distributions. The theoretical foundation for copulas is provided by Sklar's Theorem \citep{sklar1959copula}.

\begin{theorem}[Sklar’s theorem \citep{sklar1959copula}]\label{Th1}
For a $d$-dimensional random vector $X=(X_1,\ldots,X_d)$ with joint distribution function $F$ and marginal distribution functions $F_1,\ldots,F_d$, there exists a copula function $C$ such that:
\begin{equation*}
    F(x_1,...,x_d) = C[F_1(x_1),...,F_d(x_d)].
\end{equation*}
The corresponding joint density function $f$ can be expressed in terms of the copula density $c$ as:
\begin{equation*}
    f(x_1,...,x_d) = c[F_1(x_1),...,F_d(x_d)]\cdot f_1(x_1)\cdots f_d(x_d),
\end{equation*}
where $f_1,\ldots,f_d$ are the marginal densities of $F_1,\ldots, F_d$, and $c$ is the density of the copula $C$.\\
\end{theorem}

\begin{comment}
\subsubsection{Copula-based imputation}

\end{comment}

\subsection{Vine copulas}

\begin{figure}
     \centering
     \begin{subfigure}[b]{0.25\textwidth}
         \centering
         \includegraphics[width=\textwidth]{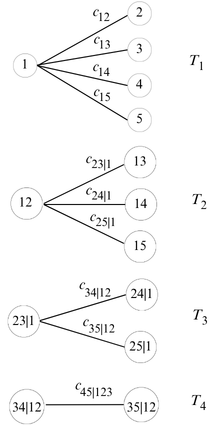}
         \caption{A C-vine}
         \label{fig:cvine}
     \end{subfigure}
     \hspace{0.05\textwidth}
     \begin{subfigure}[b]{0.45\textwidth}
         \centering
         \includegraphics[width=\textwidth]{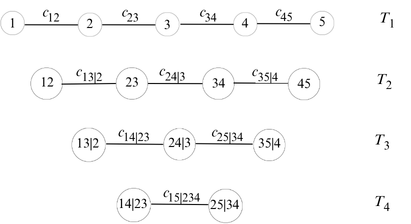}
         \caption{A D-vine}
         \label{fig:dvine}
     \end{subfigure}
        \caption{C-vine and D-vine structures with five variables}
        \label{fig:vine}
\end{figure}

A vine is a graphical framework that represents multivariate dependence through a sequence of bivariate copulas, enabling flexible modelling of pairwise dependencies among variables \citep{Bedford2002Vines, Czado2022Vine, Sahin2022Vine}. The present study uses two types of vine copulas: D-vine and C-vine. Figures~\ref{fig:cvine} and~\ref{fig:dvine} illustrate a C-vine and a D-vine structure with five variables, respectively.

In a C-vine structure, one variable acts as the central node, which links to all other variables through pairwise copulas. The C-vine copula is especially useful when a single variable plays a dominant role in capturing the dependencies among the other variables. In a D-vine structure, each variable is connected sequentially, with each variable linked to its neighbours through pairwise copulas. This structure is particularly effective when the dependencies among variables follow a chain-like or ordered pattern. As shown in Figures~\ref{fig:cvine} and~\ref{fig:dvine}, in each vine structure, nodes represent the variables (or sets of conditioned variables), while edges represent the dependencies between pairs of variables through pair copula densities. This graphical representation allows the joint density of $n$ variables to be expressed for each vine structure, as follows \citep{Aas2009Pair}: for a C- vine,
\begin{equation} \label{eq C-vine}
\begin{aligned}
    f(x_1,\ldots,x_d) &= \prod_{k=1}^{d} f_k(x_k)\times \prod_{i=1}^{d-1} c_{1,i+1} \left[F_{1}(x_1), F_{i+1}(x_{i+1})\right] \\
    &\quad \times \prod_{j=2}^{d-1} \prod_{i=1}^{d-j} c_{j,i+j|1,\ldots,j-1} \bigl[F_{j|1,\ldots,j-1}(x_j|x_1,\ldots,x_{j-1}), \cdots\\ & \cdots F_{i+j|1,\ldots,j-1}(x_{i+j}|x_1,\ldots,x_{j-1})\bigr],
\end{aligned}
\end{equation}
And for a D-vine,
\begin{equation} \label{eq D-vine}
\begin{aligned}
    f(x_1,\ldots,x_d) &= \prod_{k=1}^{d} f(x_k) \times \prod_{i=1}^{d-1} c_{i,i+1}\left[F_{i}(x_i),F_{i+1}(x_{i+1})\right] \\
    &\quad \times \prod_{j=2}^{d-1} \prod_{i=1}^{d-j} c_{i,i+j|i+1,\dots,i+j-1} \Big[F_{i|i+1,\ldots,i+j-1}(x_i|x_{i+1},\ldots,x_{i+j-1}),\cdots\\ & \cdots F_{i+j|i+1,\ldots,i+j-1}(x_{i+j}|x_{i+1},\ldots,x_{i+j-1})\Big].
\end{aligned}
\end{equation}

\noindent where the index $j$ identifies the trees while $i$ denotes the edges in each tree. In equation~\eqref{eq C-vine}, $c_{1,i+1}$, and $c_{j,i+j|1,...,j-1} $ are pair-copula densities that model the dependencies between the transformed variables $F_1(x_1)$ and $F_{i+1}(x_{i+1})$, and the conditional distributions $F(x_j|x_1,...,x_{j-1})$ and $F(x_{i+j}|x_1,...,x_{j-1})$, respectively. In equation~\eqref{eq D-vine}, $c_{i,i+1}$, and $c_{i,i+j|i+1,...,i+j-1}$ are pair-copula densities, corresponding to the dependence between the transformed variables $F_i(x_i)$ and $F_{i+1}(x_{i+1})$, and the conditional distributions $F(x_i|x_{i+1},...,x_{i+j-1})$ and $F(x_{i+j}|x_{i+1},...,x_{i+j-1})$, respectively.

\section{Proposed approach: Data fusion framework based on conditional sampling from C- or D-vines}\label{sec:pro method}

This section introduces a novel imputation-based data fusion approach based on conditional sampling from C- or D-vines. The methodology consists of two main steps. First, a C-vine or D-vine copula model is estimated to enable conditional sampling of the target variable for the donor dataset. Second, the copula parameters estimated in the first step are used to simulate the missing target variable in the recipient dataset through the conditional sampling algorithms described by \cite{Bevacqua2017Multivariate}. Section~\ref{proc: pro method} outlines the procedure for the proposed approach, and Section~\ref{sec: best vine} details the strategy for selecting the optimal vine structure for data fusion.

\subsection{Procedure for data fusion}\label{proc: pro method}

Procedure~\ref{pro:data fusion} outlines the proposed data fusion approach for imputing a single missing variable in the recipient dataset using the conditional sampling algorithms for C- and D-vines from \cite{Bevacqua2017Multivariate}. These conditional sampling algorithms allow the simulation of specific variables conditional on other variables in a sequence. 
For example, consider a dataset $\bm{X}=(\bm{x}_1,\bm{x}_2,\ldots,\bm{x}_M)$ with $M$ continuous variables and suppose a C- or D-vine structure is estimated for the variables $(U_1,..., U_m, U_{m+1},...,U_M)$, where $\bm{U}_j={F}_j(\bm{x}_j)$, $j=1,\ldots, M$. The algorithm then generates samples of $(U_{m+1},..., U_M)$ given $(U_1=u_1,..., U_m=u_m)$, where $m$ represents the number of conditioning variables. The sampling algorithms in \cite{Bevacqua2017Multivariate} assume a sequential ordering of variables for sampling. However, this ordering is not inherently present in data fusion applications, where variables are structurally missing. To address this challenge in data fusion settings involving multiple missing variables, Procedure~\ref{pro:data fusion} is extended to multivariate imputation through a series of univariate conditional sampling steps. Procedure~\ref{pro:data fusion_mult} outlines the proposed data fusion framework for multivariate imputation, illustrated using the example of two missing variables.

\begin{procedure}
\caption{Data fusion for imputing a single missing variable}
\label{pro:data fusion}
\normalsize
\begin{algorithmic}[1] % The [1] enables line numbering (optional)
    \Statex Given two independent datasets $A$ and $B$ representing same population. $A$ contains information on the set of variables, $X^A=({x}_{1}^{A},{x}_{2}^{A},\ldots,{x}_{M}^{A})$ and ${y}_{1}^{A}$ for all $n_A$ units. $B$ contains information about the set of variables, $X^B=({x}_{1}^{B},{x}_{2}^{B},\ldots,{x}_{M}^{B})$ and $Z^B=({z}_{1}^{B},{z}_{2}^{B},\ldots,{z}_{N}^{B})$ for all $n_B$ units.   
 Assume that $X^A$ and $X^B$ represent the same set of variables. The following procedure describes the imputation of ${y}_{1}^{B}$ into dataset $B$, where $A$ is the donor and $B$ the recipient:

    \begin{enumerate}
    \item[1.] \textbf{Transform to uniform scale:} Convert each observed variable in dataset $A$ and $B$ to the copula scale by applying its empirical (or assumed parametric) marginal cumulative distribution function:\linebreak ${u}_{{x}_{j}}^{A}={F}_{j}^{A}({x}_{j}^{A})$, $j=1,\ldots,M$. 

    ${u}_{y_{1}}^{A}={F}_{1}^{A}(y_{1}^A)$
    
    ${u}_{{x}_{j}}^{B}={F}_{j}^{B}({x}_{j}^{B})$, $j=1,\ldots,M$. 
    
    \item[2.] \textbf{Fit vine copula model for dataset $\bm{A}$:} Identify all candidate C-vine (or D-vine) structures for ${u}_{x_{1}}^{A},\ldots,{u}_{x_{M}}^{A},{u}_{y_{1}}^{A}$ treating (${u}_{x_{1}}^{A},\ldots, {u}_{x_{M}}^{A}$) as conditioning variables. For each vine structure, estimate a C-vine (or D-vine) copula model using the transformed data from dataset $A$ and obtain the copula parameters.

    \item[3.] \textbf{Impute missing variable in dataset $\bm{B}$:} 
    \begin{enumerate}
        \item For each vine structure, an imputed dataset for $B$ is generated using the conditional sampling algorithm for C-vine (or D-vine) of \cite{Bevacqua2017Multivariate}, with copula parameters estimated in Step 2 for the corresponding vine structure. The algorithm simulates values $\hat{u}_{y_1}^B$ from the conditional distribution of ${U_{y_1}^B}$ given observed values ${u}_{x_1}^B,\ldots, {u}_{x_M}^B$.
        \item The final imputed dataset is selected based on the optimal vine structure that best preserves the underlying dependency structure as described in Section~\ref{sec: best vine}.
    \end{enumerate}
    \item[4.] \textbf{Back-transform:} Back-transform the imputed values $\hat{u}_{y_1}^B$ to the original data scale: 
    
    $\hat{y}_{1}^B = {F^A_{y_1}}^{-1}(\hat{u}_{y_1}^B)$.  
\end{enumerate}
\end{algorithmic}
\end{procedure}

\begin{procedure}
\caption{Data fusion for imputing two missing variables}
\label{pro:data fusion_mult}
\footnotesize
\begin{algorithmic}[1] % The [1] enables line numbering (optional)
    \Statex Given two independent datasets $A$ and $B$ representing same population. $A$ contains information on the set of variables, $X^A=({x}_{1}^{A},{x}_{2}^{A},\ldots,{x}_{M}^{A})$ and $Y^A=({y}_{1}^{A},{y}_{2}^{A})$ for all $n_A$ units. $B$ contains information about the set of variables, $X^B=({x}_{1}^{B},{x}_{2}^{B},\ldots,{x}_{M}^{B})$ and $Z^B=({z}_{1}^{B},{z}_{2}^{B},\ldots,{z}_{N}^{B})$  for all $n_B$ units.   
 Assume that $X^A$ and $X^B$ represent the same set of variables. The following procedure describes the imputation of ${Y^B}=({{y}_{1}^{B}},{{y}_{2}^{B}})$ into dataset $B$, where $A$ is the donor and $B$ the recipient:

    \begin{enumerate}
    \item[1.] \textbf{Transform to uniform scale:} Convert each observed variable in dataset $A$ and $B$ to the copula scale by applying its empirical (or assumed parametric) marginal cumulative distribution function:\linebreak ${u}_{{x}_{j}}^{A}={F}_{j}^{A}({x}_{j}^{A})$, $j=1,\ldots,M$. 

    ${u}_{y_{j}}^{A}={F}_{j}^{A}(y_{j}^A)$, $j=1,2$.
    
    ${u}_{{x}_{j}}^{B}={F}_{j}^{B}({x}_{j}^{B})$, $j=1,\ldots,M$.

    \item[2. ] \textbf{Initial imputation of $y_{1}^{B}$}: Impute ${u}_{y_{1}}^{B}$ as random draws from the marginal distribution of $y_{1}^{A}$.
    
    \item[3. ] \textbf{Impute $y_{2}^{B}$ conditional on other variables:}
        \begin{enumerate}
        \item Identify all candidate C-vine (or D-vine) structures for ${u}_{x_{1}}^{A},\ldots,{u}_{x_{M}}^{A},{u}_{y_{1}}^{A},{u}_{y_{2}}^{A}$ treating (${u}_{x_{1}}^{A},\ldots, {u}_{x_{M}}^{A},{u}_{y_{1}}^{A}$) as conditioning variables. For each vine structure, estimate a C-vine (or D-vine) copula model using the transformed data from dataset $A$ and obtain the copula parameters.
        \item For each vine structure, an imputed dataset of ${u}_{y_{2}}^{B}$ is generated using the conditional sampling algorithm for C-vine (or D-vine) of \cite{Bevacqua2017Multivariate}, with copula parameters estimated in Step 3 (a). The algorithm simulates values $\hat{u}_{y_2}^B$ from the conditional distribution of ${U_{y_2}^B}$ given observed values ${u}_{x_{1}}^{B},\ldots,{u}_{x_{M}}^{B},{u}_{y_{1}}^{B}$.
        
        \item The best imputed dataset of ${u}_{y_{2}}^{B}$ is selected based on the optimal vine structure that best preserves the underlying dependency structure as described in Section \ref{sec: best vine}
        \end{enumerate}
        
    \item[4. ] \textbf{Update the imputation of $y_{1}^{B}$:}
        \begin{enumerate}
        \item Identify all candidate C-vine (or D-vine) structures for ${u}_{x_{1}}^{A},\ldots,{u}_{x_{M}}^{A},{u}_{y_{1}}^{A},{u}_{y_{2}}^{A}$ treating (${u}_{x_{1}}^{A},\ldots, {u}_{x_{M}}^{A},{u}_{y_{2}}^{A}$) as conditioning variables. For each vine structure, estimate a C-vine (or D-vine) copula model using the transformed data from dataset $A$ and obtain the copula parameters.
        \item For each vine structure, an imputed dataset of $u_{y_{1}}^{B}$ is generated using the conditional sampling algorithm for C-vine (or D-vine) of \cite{Bevacqua2017Multivariate}, with copula parameters estimated in Step 4 (b). The algorithm simulates values $\hat{u}_{y_1}^B$ from the conditional distribution of ${U_{y_1}^B}$ given observed values ${u}_{x_{1}}^{B},\ldots,{u}_{x_{M}}^{B},{u}_{y_{2}}^{B}$.
        
        \item The best imputed dataset of ${{u}_{y_{1}}^{B}}$ is selected based on the optimal vine structure that best preserves the underlying dependency structure as described in Section \ref{sec: best vine}
        \end{enumerate}

    \item[5. ] \textbf{Iterate until convergence:} Repeat Steps 3 and 4 until the imputed distribution for ${y_{1}^{B}}$ and ${y_{2}^{B}}$ stabilise (i.e., the distance between imputed and true distribution is below a defined threshold). 
    
    \item[6.] \textbf{Back-transform:} Back-transform the imputed values $\hat{u}_{y_1}^B$ and $\hat{u}_{y_2}^B$ to the corresponding original data scale:
    
    $\hat{y}_{1}^B = {F^A_{y_1}}^{-1}(\hat{u}_{y_1}^B)$.
    
    $\hat{y}_{2}^B = {F^A_{y_2}}^{-1}(\hat{u}_{y_2}^B)$.
\end{enumerate}
\end{algorithmic}
\end{procedure}

\subsection{Selection of the best vine structure}\label{sec: best vine} 

Selecting the optimal vine structure is critical to obtaining accurate fused data when using vine copulas. In data fusion, the optimal vine structure should ensure that the fused data yield a valid joint distribution of variables rather than predictions. Accordingly, the proposed approach defines the optimal vine structure as the one that best preserves dependencies among variables during data fusion. The procedure begins by imputing the missing variable in the recipient file using each candidate vine structure. Next, the average pairwise correlation differences between the donor and recipient files for each vine structure are measured using the statistic $ACD_{ext}$, as described in \ref{sec:validation}. The imputed dataset with the minimum $ACD_{ext}$ is then selected as the best imputed dataset, and the corresponding vine structure is considered the most suitable for preserving correlations during data fusion. For cases involving multiple missing variables, this procedure is applied to each variable separately in each iteration.

While this procedure enhances the preservation of the dependence structure across datasets, it becomes computationally intensive for high-dimensional datasets due to the increasing number of candidate vine structures. To improve computational efficiency, we recommend performing a preliminary analysis to identify optimal vine structures with lower $ACD_{ext}$ values using a representative subsample of the recipient file. These selected vine structures are subsequently applied to the entire dataset for imputation.

\subsection{Validation tools}\label{sec:validation}
The final step of the data fusion process involves validating the quality of the fused data. The following statistics are used to conduct a comprehensive validation of the fused data.

%use tools: ACD, ADD, MAE only (move other to the thesis)

\begin{enumerate}[label=(\Roman*)]
    
    \item Preservation of dependence structure 
     
     To assess whether imputation preserves the dependence structure across two datasets, two statistics are used.

     \begin{enumerate}[label=\arabic*.]
         \item Average correlation difference of the external coherence ($ACD_{ext}$)\citep{Aluja-Banet2013Enriching}. 
         
         $ACD_{ext}$ computes the average absolute difference of pairwise correlations between each imputed variable and the common variables computed in the donor and recipient files. $ACD_{ext}$ is defined as:

         \begin{equation*}
            ACD_{ext} = \frac{\sum^t_{j,j'} \left| {cor(\hat{y}_j^r,x_{j'}^r) - cor(y_j^d,x_{j'}^d)}\right|}{t},
        \end{equation*}

  \noindent where $cor(\hat{y}_j^r,x_{j'}^r)$ and $cor(y_j^d,x_{j'}^d)$ refer to the pairwise correlation between the imputed variable $y$ and each common variable $x$ in the recipient file ($r$) and donor file ($d$), respectively. $t$ is the total number of pairs of variables $(j,j')$.

         \item Average correlation difference of the internal coherence ($ACD_{int}$)\citep{Aluja-Banet2013Enriching}.

         $ACD_{int}$ computes the average absolute difference of correlations computed between the imputed variables in the donor and recipient files. $ACD_{int}$ is defined as:
     
         \begin{equation*}
            ACD_{int} = \frac{\sum^s_{j,j'} \left| {cor(\hat{y}_j^r,\hat{y}_{j'}^r) - cor(y_j^d,y_{j'}^d)}\right|}{s},
        \end{equation*}
        
        \noindent where $cor(\hat{y}_j^r,\hat{y}_{j'}^r)$ and $cor(y_j^d,y_{j'}^d)$ refer to the pairwise correlation between the imputed variables in the recipient file ($r$) and the corresponding observed variables in the donor file ($d$), respectively. $s$ is the total number of pairs of imputed variables$(j,j')$. 

     \end{enumerate}

Lower values of $ACD_{ext}$ and $ACD_{int}$ indicate better preservation of the underlying dependence structure.

    \item Preservation of imputed distributions

     To assess the preservation of the marginal distributions of imputed variables, the statistical distance between the probability distributions of each imputed variable in the recipient file and the corresponding observed variable in the donor file is calculated using Kullback–Leibler (KL) divergence. Distance measured using KL is defined as:

      \begin{equation*}
    D_{\mathrm{KL}}= \sum_{x \in \mathcal{X}} P(x)\, \log \left( \frac{P(x)}{Q(x)} \right), 
        \end{equation*}
     where $Q(x)$ is the imputed distribution in the recipient file and $P(x)$ is the corresponding true distribution in the donor file. 
     A lower value of $D_{\mathrm{KL}}$ indicates better preservation of the imputed distribution.
     
\end{enumerate}

\section{Simulation study }
\begin{figure}[h]
    \centering
    \includegraphics[scale=0.6]{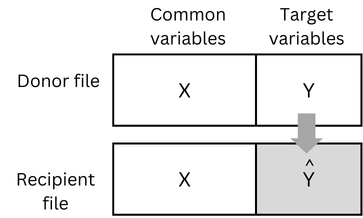}
    \caption{Data fusion diagram for simulation study}
    \label{fig: sim_df diagram}
\end{figure}

This section evaluates the proposed data fusion approach through a simulation study using the Australian Bureau of Statistics (ABS) Survey of Income and Housing 2019-20 data. For the study, a subset of the dataset is constructed by selecting five variables: working hours ($x_1$), disposable income ($x_2$), equivalised net wealth ($x_3$), employee income ($y_1$), and investment income ($y_2$) and restricting the sample to positive employee income, resulting in 15,843 observations. Two independent samples, each consisting of 2,500 observations, are randomly selected from this subset. One sample is designated as the donor and the other as the recipient. The data simulation is repeated 100 times to account for sampling variability. The variables $Y=(y_1, y_2)$ are removed from only the recipient file, and the common set of variables for both donor and recipient files is defined as $X = (x_1, x_2, x_3)$. Two missing variables, $y_1$ and $y_2$, are then imputed into the recipient file within the data fusion framework, as illustrated in Figure~\ref{fig: sim_df diagram}.

Two proposed data fusion methods based on C-vine (CVIMP) and D-vine (DVIMP) are implemented. Their performance is evaluated relative to the two comparative methods, PMM and RF, using the validation tools described in Section~\ref{sec:validation}. PMM is selected as a comparative method because it is widely used and well-established as a standard for multiple imputation, serving as a reliable benchmark for performance evaluation. RF is also chosen, as \cite{Emmenegger2023Evaluating} demonstrated its superior performance compared to other tree-based machine learning approaches, such as CART (implemented via Rpart in R), in a data fusion application. The robustness of RF and its capacity to model complex relationships further justify its use as a benchmark. PMM and RF are implemented in R using the \texttt{mice} package. Multiple imputation is performed for all methods, and validation criteria are averaged across 10 imputed datasets to produce a single estimate per simulation run. The results are described in Section~\ref{sec: sim result}. 

\subsection{Results}\label{sec: sim result}

Table~\ref{tab:sim_ACD} presents $ACD_{ext_{x,y_1}}$ and $ACD_{ext_{x,y_2}}$, which assess the preservation of correlations between common variables ($X$) and each imputed variable, $y_1$ and $y_2$, respectively. The table also reports $ACD_{int_{y1, y2}}$, which evaluates the preservation of correlations between $y_1$ and $y_2$. Results indicate that DVIMP and CVIMP yield lower values for both $ACD_{ext}$ and $ACD_{int}$ than alternatives, demonstrating the better preservation of the correlation structure between each imputed and common variables, as well as among the imputed variables across two datasets. 
%These results suggest that the proposed methods most accurately reconstruct the multivariate distribution compared to alternative approaches.

Table~\ref{tab: sim KL} presents $D_{{\mathrm{KL}}_{y_1}}$ and $D_{{\mathrm{KL}}_{y_2}}$, which assess the preservation of the marginal distributions of the imputed variables $y_1$ and $y_2$, respectively. The $D_{{\mathrm{KL}}_{y_1y_2}}$ metric evaluates the preservation of the multivariate distribution of ($y_1, y_2$), estimated using two-dimensional kernel density estimation. The results show that both DVIMP and CVIMP yield lower distance values for $D_{{\mathrm{KL}}_{y_1}}$ and $D_{{\mathrm{KL}}_{y_2}}$, indicating more accurate reproduction of the individual distributions of each imputed variable compared to other methods. Similarly, the lower values of $D_{{\mathrm{KL}}_{y_1y_2}}$ obtained by DVIMP and CVIMP suggest that these methods more accurately preserve the multivariate distribution of the imputed variables across two datasets than comparative methods. 
%Overall, these findings indicate that the proposed methods surpass comparative approaches in preserving both the marginal and joint distributions of the imputed variables across two datasets.

\setlength{\tabcolsep}{8pt}
\begin{table}[ht!]
\centering
\caption{Validation results for $ACD_{ext}$ and $ACD_{int}$ in simulation study. Values are mean (first line) and (2.5\%, 97.5\%) quantiles (second line) across 100 simulation runs.}
\label{tab:sim_ACD}
\begin{tabular}{lccc}
\toprule
\textbf{Method} & \textbf{$ACD_{ext_{x,y_1}}$} & \textbf{$ACD_{ext_{x,y_2}}$} & \textbf{$ACD_{int_{y1,y2}}$}  \\  
\midrule

DVIMP &
\begin{tabular}[t]{@{}c@{}}
\textbf{0.012} \\
\textbf{(0.006, 0.022)}
\end{tabular}
&
\begin{tabular}[t]{@{}c@{}}
\textbf{0.016} \\
\textbf{(0.009, 0.029)}
\end{tabular}
&
\begin{tabular}[t]{@{}c@{}}
\textbf{0.017}\\
\textbf{(0.009, 0.028)}
\end{tabular}
\\
\addlinespace
CVIMP &
\begin{tabular}[t]{@{}c@{}}
\textbf{0.017} \\
\textbf{(0.007, 0.029)}
\end{tabular}
&
\begin{tabular}[t]{@{}c@{}}
\textbf{0.017} \\
\textbf{(0.010, 0.024)}
\end{tabular}
&
\begin{tabular}[t]{@{}c@{}}
\textbf{0.018} \\
\textbf{(0.008, 0.031)}
\end{tabular}
\\
\addlinespace
RF &
\begin{tabular}[t]{@{}c@{}}
0.020 \\
(0.012, 0.044)
\end{tabular}
&
\begin{tabular}[t]{@{}c@{}}
0.023 \\
(0.014, 0.043)
\end{tabular}
&
\begin{tabular}[t]{@{}c@{}}
0.031 \\
(0.013, 0.068)
\end{tabular}
\\
\addlinespace
PMM &
\begin{tabular}[t]{@{}c@{}}
0.093 \\
(0.040 0.138)
\end{tabular}
&
\begin{tabular}[t]{@{}c@{}}
0.086 \\
(0.061, 0.114)
\end{tabular}
&
\begin{tabular}[t]{@{}c@{}}
0.168 \\
(0.089, 0.289)
\end{tabular}
\\

\bottomrule
\end{tabular}
\end{table}

\setlength{\tabcolsep}{8pt}
\begin{table}[ht!]
\centering
\caption{Validation results for $D_{\mathrm{KL}}$ in simulation study. Values are mean (first line) and (2.5\%, 97.5\%) quantiles (second line) across 100 simulation runs.}
\label{tab: sim KL}
\begin{tabular}{lccc}
\toprule
\textbf{Method} & \textbf{$D_{{\mathrm{KL}}_{y_1}}$} & \textbf{$D_{{\mathrm{KL}}_{y_1}}$} & \textbf{$D_{{\mathrm{KL}}_{y_1y_2}}$} \\ 
\midrule

DVIMP &
\begin{tabular}[t]{@{}c@{}}
\textbf{0.023} \\
\textbf{(0.006, 0.068)}
\end{tabular}
&
\begin{tabular}[t]{@{}c@{}}
\textbf{0.032} \\
\textbf{(0.003, 0.134)}
\end{tabular}
&
\begin{tabular}[t]{@{}c@{}}
\textbf{0.045} \\
\textbf{(0.023, 0.067)}
\end{tabular}
\\
\addlinespace
CVIMP &
\begin{tabular}[t]{@{}c@{}}
\textbf{0.018} \\
\textbf{(0.005, 0.058)}
\end{tabular}
&
\begin{tabular}[t]{@{}c@{}}
 \textbf{0.025}\\
\textbf{(0.003, 0.086)}
\end{tabular}
&
\begin{tabular}[t]{@{}c@{}}
 \textbf{0.047}\\
\textbf{(0.023, 0.077)}
\end{tabular}
\\
\addlinespace
RF &
\begin{tabular}[t]{@{}c@{}}
0.034 \\
(0.012, 0.094)
\end{tabular}
&
\begin{tabular}[t]{@{}c@{}}
0.048 \\
(0.007, 0.170)
\end{tabular}
&
\begin{tabular}[t]{@{}c@{}}
0.060 \\
(0.034, 0.115)
\end{tabular}
\\
\addlinespace
PMM &
\begin{tabular}[t]{@{}c@{}}
0.079 \\
(0.015, 0.299)
\end{tabular}
&
\begin{tabular}[t]{@{}c@{}}
0.187 \\
(0.028, 0.875)
\end{tabular}
&
\begin{tabular}[t]{@{}c@{}}
0.118 \\
(0.049, 0.234)
\end{tabular}
\\

\bottomrule
\end{tabular}
\end{table}

\section{Application of data fusion: Estimating negative income distribution in survey data using administrative tax data}

This section validates the proposed data fusion approach in a real-world setting, estimating income statistics in the household income survey, incorporating information from administrative tax records. The performance of the proposed method is compared with alternative approaches used in the simulation study.

Negative income reported in survey data, which may result from small-business losses, investment write-downs, or self-employment, serves as a key indicator of financial volatility, income risk, and economic vulnerability \citep{Julian2022Neginc}. Nevertheless, negative income values are frequently under-reported, misclassified, or truncated in survey responses. For instance, respondents may inaccurately recall net losses or report zero income instead of a negative value \citep{Angel2019Neginc, Hlasny2022Neginc}. This under-reporting can systematically underestimate income variability and introduce bias into measures of inequality and poverty \citep{Hlasny2022Neginc}. Furthermore, negative income reflects exposure to income shocks, a critical aspect of economic security that has become increasingly relevant in the contemporary gig economy and among small enterprises \citep{Bufe2021Neginc,oecd2023shaky}. Administrative data, such as records from the Australian Taxation Office (ATO), generally provide a more accurate account of these losses through tax return filings and thus offer a more reliable benchmark for the actual income distribution. Aligning survey-based income distributions with administrative benchmarks improves the representativeness and accuracy of derived indicators, including measures of inequality, income mobility, and comparisons between taxable and disposable income \citep{Flachaire2023Neginc}. Accordingly, this study applies the proposed vine copula-based data fusion approach to estimate the negative income distribution in survey data, using administrative records as a benchmark to obtain estimates that more closely align with the true distribution observed in administrative data. 

\subsection{Data}
This study uses two data sources: the ABS Survey of Income and Housing (SIH) and the ABS Multi-Agency Data Integration Project (MADIP) for the 2019-20 financial year.

The 2019-20 SIH collected detailed information on income, wealth, housing, household characteristics, and personal characteristics from a sample of 15,011 Australian households between July 2019 and June 2020. The SIH enables analysis and monitoring of the social and economic welfare of Australian residents in private dwellings. The MADIP dataset, renamed the Person Level Integrated Data Asset (PLIDA) in 2023, is a secure data asset that integrates administrative data on health, education, government payments, income and taxation, employment, and population demographics, including Census data, over time. It provides comprehensive insights into various population groups in Australia. This study uses the ATO data records for 2019-20, linked to the 2021 Census from the MADIP database. 

In the analysis, the negative total income across both datasets (SIH and ATO) is considered as the target variable. The SIH dataset identifies 206 individuals with negative income, while the ATO dataset includes 65,990 such individuals (a subset of available ATO records following data cleaning and preparation). Table~\ref{tab:compar_sum} presents summary statistics for the negative total income variable across both datasets. Although SIH and ATO distributions exhibit negative values for skewness, mean, median, and quartiles, the SIH distribution values have smaller magnitudes than the ATO values. A bootstrap analysis indicated that the differences between the two sources are not just due to sampling variability. 

Overall, Table~\ref{tab:compar_sum} demonstrate that SIH may underestimate the magnitude of negative income distribution compared to the ATO data, although the definitions of total income in SIH and ATO can differ. To evaluate the proposed data fusion approach in a real-data setting, the ATO data, as a more comprehensive source of negative income information, are considered the benchmark for the true income distribution in this context. The proposed approach is then applied to estimate a negative income distribution for the survey data that more closely aligns with the ATO distribution.

\begin{table}[ht!]
\centering
\caption{ Summary statistics of the negative total income data from SIH and from ATO datasets. For ATO data, values in parentheses represent the 95\% bootstrap confidence intervals estimated from 1000 subsamples of size 206 to match the SIH sample size.}
\label{tab:compar_sum}
\begin{tabular}{lcc}
\toprule
Statistic & ATO & SIH \\
\midrule

25th percentile &
-716.37 (-962.45, -515.71)
& -337.41 \\

\addlinespace
Median &
-238.26 (-315.85, -179.16)
& -128.97 \\

\addlinespace
75th percentile &
-76.60 (-103.38, -54.37)
&  -52.78 \\

\addlinespace
Mean &
-801.91 (-1033.98, -579.97) 
& -359.77 \\

\addlinespace
Skewness &
-5.77 (-7.66, -2.97)
& -4.99 \\

\bottomrule
\end{tabular}
\end{table}

\subsubsection{Matching variables} \label{matching}
Selecting appropriate matching variables is essential in data fusion, as the validity of the matching process depends on the comparability of common variables across datasets. In this study, the matching variables are selected based on the approximate similarity of their marginal distributions across datasets. As a result, two income-related variables, unincorporated business income and investment income, are selected, as both contribute to the construction of total income in each dataset and capture relevant aspects of individuals’ income composition. In addition, two demographic variables, age and highest level of education, are included to capture important socioeconomic characteristics. 

Due to the limited availability of common variables across datasets, all selected variables that meet the specified criteria are retained to facilitate matching and maintain consistency across sources. This approach reflects typical real-world data-fusion scenarios, where common variables are often scarce. When a broader set of common variables is available, using formal variable selection techniques can enhance matching quality by identifying the most informative variables.

\subsubsection{Harmonising two datasets}

Before undertaking data fusion, it is essential to establish a common ground between the data sources. The two datasets in this study (SIH and ATO data) are already comparable in terms of reporting unit and reference period, as both are collected at the person level and correspond to the same financial year. This section further ensures comparability of the two datasets in terms of population coverage and representativeness, and describes the harmonisation of matching variables.

Regarding population coverage, the SIH represents a sample of both taxpayers and non-taxpayers in Australia, whereas the ATO data only covers individuals within the tax system. 
In this study, the analysis is restricted to individuals with negative total income in both datasets. Individuals in this group typically have no taxable income and therefore no tax payable, providing a common economic characteristic of the two sources. Restricting the analysis to this subgroup improves comparability between the two datasets with respect to income composition and tax status. However, it is important to acknowledge that the two datasets are not perfectly comparable; some compositional differences remain. In practice, data fusion rarely involves perfectly aligned populations, and the objective is to reconcile two imperfect but complementary data sources to produce improved estimates. Therefore, while acknowledging the inherent differences among target populations, the imposed restriction provides a reasonable basis for using ATO data as a benchmark to estimate negative income in the SIH data. This supports the evaluation of the proposed approach in this context.

Among matching variables, the definitions of two income-related variables are broadly consistent across the datasets, although their reporting scales differ. The ATO provides annual income values, whereas the SIH reports weekly income. To ensure comparability, ATO annual values are converted to weekly amounts by dividing by 52. The two data sources also classify educational level differently. This variable is harmonised by re-categorising the original classifications so that both datasets use the same categories, thereby ensuring comparability. Since educational level is measured as an ordinal categorical variable and the proposed approach is designed for continuous variables, this variable is transformed using its empirical cumulative distribution function to enable inclusion in the analysis.

\subsection{Data fusion implementation}\label{preparation}

The analysis includes records for individuals aged 15 years and older who report negative total income in both data sources. The SIH dataset is complete with no missing values while the ATO records exhibit missing data in the common variables. Incomplete records are excluded, resulting in 65,990 ATO records and 206 SIH records available for data fusion. Two proposed methods, DVIMP and CVIMP, and two comparative methods, PMM and RF, are implemented to impute negative total income data to the SIH dataset. Results are discussed in the section~\ref{sec: sim_Results} 

\subsubsection{Results} \label{sec: sim_Results} 

Table~\ref{tab:real_val} indicates that $ACD_{ext}$ values are broadly similar across the proposed methods (DVIMP and CVIMP) and RF. However, the proposed vine copula-based methods yield smaller values, suggesting improved preservation of the correlation structure between imputed and observed variables. While all methods exhibit reasonably low $D_{\mathrm{KL}}$ values, the proposed vine copula-based methods achieve lower values than the alternatives, indicating better preservation of the imputed distribution. The summary statistics in Table~\ref{tab:real_summary} show that DVIMP and CVIMP generate values closer to the true negative income distribution in the donor, with DVIMP providing the closest alignment. In contrast, RF and PMM exhibit considerable discrepancies in summary measures relative to the true distribution, indicating that these methods do not reproduce key distributional characteristics as accurately as the proposed vine copula-based approaches. Collectively, these results provide evidence that the proposed methods more accurately estimate the negative income distribution in survey data while preserving the underlying multivariate dependence structure across two datasets (ATO and SIH).

\begin{table}[ht!]
\centering
\caption{Validation results for $ACD_{ext}$ and $D_{KL}$ in real data application. Values are presented as means across 10 imputed datasets.}
\label{tab:real_val}
\begin{tabular}{p{3cm}p{3.3cm}p{3cm}}
\toprule
Method & {$ACD_{ext}$} (std. err) & {$D_{\mathrm{KL}}$} (std. err) \\  
\midrule
DVIMP & \textbf{0.154} (0.006)  & \textbf{0.031} (0.004) \\
CVIMP & \textbf{0.170} (0.006) & \textbf{0.037} (0.004)\\
RF & 0.194 (0.006) & 0.053 (0.008)\\
PMM & 0.233 (0.004) & 0.062 (0.008) \\
\bottomrule
\end{tabular}
\end{table}

\begin{table}[ht!]
\centering
\caption{Comparison of summary statistics for true negative income in the donor file (ATO data) and estimated negative income for the recipient file (SIH data) obtained using different data fusion methods.}
\label{tab:real_summary}
\begin{tabular}{lrrrr}
\toprule
Method & 25th percentile & Median & 75th percentile & Mean \\  
\midrule
True (ATO) & \textbf{-716.38}  & \textbf{-238.30} & \textbf{-76.60} & \textbf{-801.91}\\
Estimated - DVIMP & \textbf{-599.13} & \textbf{-227.04} & \textbf{-97.47} & \textbf{-760.31} \\
Estimated - CVIMP & \textbf{-587.13} & \textbf{-212.96} & \textbf{-104.23} & \textbf{-653.05} \\
Estimated - RF & -332.97 & -130.66 &-58.00 & -360.46 \\
Estimated - PMM & -317.03 & -111.16 &  -38.90 & -333.09 \\
\bottomrule
\end{tabular}
\end{table}

\section{Discussion}

This paper proposes a novel imputation-based data fusion approach using two types of vine copulas, C-vines and D-vines, that provides a flexible framework to improve the quality of the integrated data by preserving the marginal distributions of the imputed variables and the underlying multivariate dependence structure across datasets. The proposed approach uses conditional sampling from C-vine and D-vine copulas, introduced by \cite{Bevacqua2017Multivariate}, as the imputation mechanism. To the best of our knowledge, this is the first study to implement a vine copula-based imputation within a data fusion context.

The proposed vine copula-based data fusion approach offers several methodological advantages. First, vine copula models provide a flexible framework for modelling complex multivariate dependence structures, allowing different copula families to capture non-linear associations, asymmetric relationships, and tail dependence beyond simple linear correlation. Second, the approach generates multiple imputed datasets using candidate vine structures and selects the optimal structure based on its ability to preserve the dependence structure, as assessed through correlation-based validation metrics. This selection mechanism ensures that the final imputed dataset maintains the multivariate relationships observed in the donor data as closely as possible. Third, imputation is performed on the copula scale and subsequently back-transformed to the original scale using the empirical marginal distributions from the donor data. This step ensures that the imputed values follow the same marginal distribution as the observed donor data, preventing unrealistic values. Consequently, these features enable the proposed method to produce higher-quality fused data by preserving both imputed distributions and multivariate dependence structure. Additionally, existing data fusion methods typically employ standard imputation techniques that primarily aim to minimise prediction error when imputing missing values within a single dataset. In contrast, the proposed approach focuses on imputing missing variables across multiple datasets to accurately reconstruct the joint distribution of the observed and imputed variables, rather than optimising prediction accuracy. This strategy enhances the reliability of downstream analyses that depend on valid multivariate relationships. 

The results from both the simulation study and the real-data application show that the proposed approaches based on C-vine and D-vine structures improve the quality of fused data by preserving imputed distributions and the multivariate dependence structure compared with alternative methods. In the real-data application, both D-vine and C-vine methods demonstrate comparable performance; however, the D-vine approach yields a slight improvement in estimating the negative income distribution. These results indicate that the D-vine model more effectively captures the underlying dependence structure in the data compared to the C-vine model.

Future research will extend the proposed approach to accommodate mixed data types, including discrete, categorical, and continuous variables, using a latent variable framework as implemented in previous work for modelling copulas for mixed data \citep{Jiryaie12016Latent, Tekumalla2017Vine, Lu2025Latent}. The current study validates the proposed method under the data fusion assumption that the conditional independence assumption (CIA) holds \citep{Rodgers1984Evaluation}. In practical scenarios where the CIA is violated, future work will extend the proposed method to relax this assumption, thereby reducing inference bias arising from such violations.

\section*{Conflict of interest}

The authors declared no potential conflicts of interest concerning the research, authorship, and/or publication of this article.

\section*{Acknowledgements}

The authors gratefully acknowledge support from the Australian Bureau of Statistics under grant number ABS2020.325. The authors also thank Ruel Abello from the Australian Bureau of Statistics for valuable feedback and advice on this work.

\section*{Data availability statement}
The data used in this study are not publicly available due to confidentiality restrictions. Access to the data was provided through the Australian Bureau of Statistics (ABS) DataLab. Researchers may apply for access through the ABS, subject to approval and data access conditions.
 
 % or abbrv, unsrt, alpha
\bibliography{references}

@article{Aas2009Pair,
title = {Pair-copula constructions of multiple dependence},
journal = {Insurance: Mathematics and Economics},
volume = {44},
number = {2},
pages = {182-198},
year = {2009},
author = {Kjersti Aas and Claudia Czado and Arnoldo Frigessi and Henrik Bakken},
doi = {10.1016/j.insmatheco.2007.02.001}
}

@Inbook{Aluja-Banet2013Enriching,
author="Tom Aluja-Banet
and Josep Daunis-i-Estadella 
and Yan Hong Chen",
title="Enriching a Large-Scale Survey from a Representative Sample by Data Fusion: Models and Validation",
bookTitle="Survey Data Collection and Integration",
year="2013",
publisher="Springer Berlin Heidelberg",
pages="121-137"
}

@article{Angel2019Neginc,
    author = {Angel, Stefan and Disslbacher, Franziska and Humer, Stefan and Schnetzer, Matthias},
    title = {What did you Really Earn Last Year?: Explaining Measurement Error in Survey Income Data},
    journal = {Journal of the Royal Statistical Society Series A: Statistics in Society},
    volume = {182},
    number = {4},
    pages = {1411-1437},
    year = {2019},
    month = {04},
    issn = {0964-1998},
    doi = {10.1111/rssa.12463},
    url = {https://doi.org/10.1111/rssa.12463}
}

@article{Bedford2002Vines,
  author = {Tim Bedford and Roger M. Cooke},
 journal = {The Annals of Statistics},
 number = {4},
 pages = {1031--1068},
 publisher = {Institute of Mathematical Statistics},
 title = {Vines: A new graphical model for dependent random variables},
 volume = {30},
 year = {2002},
doi={10.1214/aos/1031689016}
}

@Article{Bevacqua2017Multivariate,
AUTHOR = {Bevacqua, E. and Maraun, D. and Hob{\ae}k Haff, I. and Widmann, M. and Vrac, M.},
TITLE = {Multivariate statistical modelling of compound events via pair-copula constructions: analysis of floods in Ravenna (Italy)},
JOURNAL = {Hydrology and Earth System Sciences},
VOLUME = {21},
YEAR = {2017},
NUMBER = {6},
PAGES = {2701--2723},
DOI = {10.5194/hess-21-2701-2017}
}

@article{Bufe2021Neginc,
author = {Bufe, Sam and Roll, Stephen and Kondratjeva, Olga and Skees, Stephanie and Grinstein-Weiss, Michal},
year = {2021},
month = {10},
pages = {379-407},
title = {Financial Shocks and Financial Well-Being: What Builds Resiliency in Lower-Income Households?},
volume = {161},
journal = {Social Indicators Research},
doi = {10.1007/s11205-021-02828-y}
}

@article{Chapon2023Imp,
title = {Imputation of missing values in environmental time series by D-vine copulas},
journal = {Weather and Climate Extremes},
volume = {41},
pages = {100591},
year = {2023},
issn = {2212-0947},
doi = {10.1016/j.wace.2023.100591},
author = {Antoine Chapon and Taha B.M.J. Ouarda and Yasser Hamdi}
}

@article{Chiara2019Integrated,
author = {Elena Dalla Chiara and Martina Menon and Federico Perali},
title = {An Integrated Database to Measure Living Standards},
journal = {Journal of Official Statistics},
number = {3},
volume = {35},
year = {2019},
pages = {531--576}
}

@article{Czado2022Vine,
  author  = {Czado, Claudia and Nagler, Thomas},
  title   = {Vine Copula Based Modeling},
  journal = {Annual Review of Statistics and Its Application},
  year    = {2022},
  volume  = {9},
  pages   = {453--477},
  doi     = {10.1146/annurev-statistics-040220-101153}
}

@article{Emmenegger2023Evaluating,
    author = {Jana Emmenegger and Ralf Münnich and Jannik Schaller},
    title = {Evaluating Data Fusion Methods to Improve Income Modeling},
    journal = {Journal of Survey Statistics and Methodology},
    volume = {11},
    number = {3},
    pages = {643-667},
    year = {2023},
    publisher={Oxford University Press}
}

@ARTICLE{Ene2013CVine,
	author = {Ene, Liviu Theodor and Næsset, Erik and Gobakken, Terje},
	title = {Model-based inference for $k$-nearest neighbours predictions using a canonical vine copula},
	year = {2013},
	journal = {Scandinavian Journal of Forest Research},
	volume = {28},
	number = {3},
	pages = {266 – 281},
	doi = {10.1080/02827581.2012.723743}
}

@article{Flachaire2023Neginc,
author = {Flachaire, Emmanuel and Lustig, Nora and Vigorito, Andrea},
title = {Underreporting of Top Incomes and Inequality: A Comparison of Correction Methods using Simulations and Linked Survey and Tax Data},
journal = {Review of Income and Wealth},
volume = {69},
number = {4},
pages = {1033-1059},
doi = {https://doi.org/10.1111/roiw.12618},
url = {https://onlinelibrary.wiley.com/doi/abs/10.1111/roiw.12618},
year = {2023}
}

@article{Fosdick2016Categorical,
author = { B. K. Fosdick AND M. DeYoreo AND J. P. Reiter},
title ={Categorical Data Fusion Using Auxiliary Information},
journal = {The Annals of Applied Statistics},
volume = {10},
number = {4},
pages = {1907-1929},
year = {2016}
}

@article{Hasler2018Vine,
    title = {Vine Copulas for Imputation of Monotone Non-response},
    author = {Caren Hasler and  Radu V. Craiu and Louis-Paul Rivest},
    journal = {International Statistical Review},
    volume = {86},
    number = {3},
    pages = {488-511},
    year = {2018}
}

@article{Hlasny2022Neginc,
author = {Hlasny, Vladimir and Ceriani, Lidia and Verme, Paolo},
title = {Bottom Incomes and the Measurement of Poverty and Inequality},
journal = {Review of Income and Wealth},
 year = {2021},
volume = {68},
number = {4},
pages = {970-1006},
doi = {https://doi.org/10.1111/roiw.12535},
url = {https://onlinelibrary.wiley.com/doi/abs/10.1111/roiw.12535}
}

@article{Jiryaie12016Latent,
author = {F. Jiryaie and N. Withanage and B. Wu and A.R. de Leon},
title = {Gaussian copula distributions for mixed data, with application in discrimination},
journal = {Journal of Statistical Computation and Simulation},
volume = {86},
number = {9},
pages = {1643--1659},
year = {2016},
publisher = {Taylor \& Francis},
doi = {10.1080/00949655.2015.1077386},
URL = { https://doi.org/10.1080/00949655.2015.1077386}
}

@article{Julian2022Neginc,
title = {The effect of negative income shocks on pensioners},
journal = {Labour Economics},
volume = {76},
pages = {102175},
year = {2022},
issn = {0927-5371},
doi = {https://doi.org/10.1016/j.labeco.2022.102175},
url = {https://www.sciencedirect.com/science/article/pii/S0927537122000665},
author = {Julian V. Johnsen and Alexander Willén}
}

@article{Lu2025Latent,
title = {Toward accurate credit evaluation: an efficient imputation approach for financial data},
journal = {Data Science and Management},
volume = {8},
number = {3},
pages = {374-387},
year = {2025},
issn = {2666-7649},
doi = {https://doi.org/10.1016/j.dsm.2025.06.001},
url = {https://www.sciencedirect.com/science/article/pii/S2666764925000281},
author = {Jie Lu and Shengda Zhuo and Jinjie Qiu and Yin Tang}
}

@article{Moretti2023Improving,
    author = {Angelo Moretti and Natalie Shlomo},
    title = "{Improving Statistical Matching when Auxiliary Information is Available}",
    journal = {Journal of Survey Statistics and Methodology},
    volume = {11},
    number = {3},
    pages = {619-642},
    year = {2023},
    publisher = {Oxford University Press}
}

@book{oecd2023shaky,
  title        = {On Shaky Ground? Income Instability and Economic Insecurity in Europe},
  author       = {{OECD}},
  year         = {2023},
  publisher    = {OECD Publishing},
  address      = {Paris},
  doi          = {10.1787/9bffeba6-en},
  url          = {https://doi.org/10.1787/9bffeba6-en}
}

@article{Rodgers1984Evaluation,
    author = {Willard L. Rodgers},
    title = {An Evaluation of Statistical Matching},
    journal = {Journal of Business \& Economic Statistics},
    volume = {2},
    number = {1},
    pages = {91–102},
    year = {1984}
}

@inproceedings{sklar1959copula,
  title={Fonctions de r{\'e}partition {\`a} n dimensions et leurs marges},
  author={Sklar, M},
  booktitle={Annales de l'Institut de Statistique de l'Université de Paris},
  volume={8-3},
  pages={229--231},
  year={1959}
}

@article{Tekumalla2017Vine,
    title = {Vine copulas for mixed data : multi-view clustering for mixed data beyond meta-Gaussian dependencies},
    author = {Lavanya Sita Tekumalla and   Vaibhav Rajan and Chiranjib Bhattacharyya },
    journal = {Machine Learning},
    volume = {106},
    pages = {1331–1357},
    year = {2017},
    publisher = {Springer}
}

@article{Sahin2022Vine,
title = {Vine copula mixture models and clustering for non-Gaussian data},
journal = {Econometrics and Statistics},
volume = {22},
pages = {136-158},
year = {2022},
note = {The 2nd Special issue on Mixture Models},
issn = {2452-3062},
doi = {https://doi.org/10.1016/j.ecosta.2021.08.011},
author = {Özge Sahin and Claudia Czado}
}

\end{document}